\documentclass[12pt,nofootinbib]{article}

\usepackage[dvips]{graphicx}
\usepackage{amssymb}

\usepackage{amsmath}
\usepackage{epsfig}

\def\beq{\begin{equation}}
\def\eeq{\end{equation}}
\def\be{\begin{equation}}
\def\ee{\end{equation}}
\def\bea{\begin{eqnarray}}
\def\eea{\end{eqnarray}}

\DeclareRobustCommand{\SkipTocEntry}[4]{}

\begin{document}

\begin{titlepage}

\setcounter{page}{1} \baselineskip=15.5pt \thispagestyle{empty}

\begin{flushright}
hep-th/0703250\\
PUPT-2229
\end{flushright}
\vfil

\begin{center}
{\LARGE Primordial Black Hole Baryogenesis}
\end{center}
\bigskip\

\begin{center}
{\large Daniel Baumann$^{1}$, Paul Steinhardt$^{1,2}$, and Neil Turok$^{3}$}
\end{center}

\begin{center}
\textit{$^{1}$Department of Physics, Jadwin Hall,
Princeton University, Princeton, NJ 08544}\\
\textit{$^{2}$Princeton Center for Theoretical Physics,
Princeton University, Princeton, NJ 08544}\\
\textit{$^{3}$DAMTP,
CMS, Wilberforce Road, Cambridge, CB3\,0WA, UK}
 \end{center} \vfil

\noindent
We reconsider the possibility that the observed baryon asymmetry was
generated by the evaporation
of primordial black holes that dominated the
early universe. We present a simple derivation showing that the
baryon asymmetry is insensitive to the initial black hole density
and the cosmological model but is sensitive to the temperature-dependence of
the CP and baryon-violating (or lepton-violating) interactions.
We also consider the possibility that black holes stop evaporating and
form Planck-mass remnants that act as dark matter. 
We show that  primordial
black holes cannot simultaneously account for both
the observed baryon asymmetry and the (remnant) dark matter density unless the magnitude of CP violation is much greater than expected from most particle physics models.
Finally, we apply these results to ekpyrotic/cyclic models, in which primordial black holes may form 
when branes collide. 
We find that obtaining the observed baryon asymmetry 
is compatible with the other known constraints on parameters.

\vfil
\begin{flushleft}
\today
\end{flushleft}

\end{titlepage}

\newpage

\section{Introduction}
\label{sec:intro}
Evidence for matter--anti-matter asymmetry is inferred from 
studies of the particle content in cosmic rays, the absence of an intense $\gamma$-ray background from matter--anti-matter 
annihilations on cluster scales \cite{Cohen:1997ac}, big-bang nucleosynthesis (BBN)~\cite{BBN} and high 
precision measurements of the cosmic microwave background (CMB)~\cite{Spergel:2003cb}. However, imposing 
the observed baryon-to-entropy value, $B =  \frac{n_B}{s} = \frac{n_b-n_{\bar{{b}}}}{s} = 9.2^{+0.6}_{-0.4} \times 10^{-11}$, 
as an initial condition seems extremely unnatural.
The currently favored alternative is to imagine that the baryon asymmetry was 
generated from symmetric initial conditions through a dynamical
mechanism.

The idea of producing an excess of baryons over anti-baryons in black hole evaporation goes back to
Hawking~\cite{Hawking} and Zeldovich~\cite{Zeldovich}. 
More recently, this notion has been explored extensively in the context of grand unified theories (GUTs) 
({\it e.g.}~\cite{earlypapers, barrow, therest}). 
As a black hole evaporates, its horizon temperature eventually becomes higher than the GUT-scale.
Hawking evaporation of a gas of primordial black holes provides out-of-equilibrium conditions,
whereas grand unified theories generically violate CP and baryon number conservation. 
In particular, evaporating black holes may emit massive particles whose decay generates a net baryon asymmetry.
Hence,
black hole evaporation and local GUT-scale physics together satisfy the {\it Sakharov 
conditions}~\cite{Sakharov:dj} and may
provide a natural mechanism for generating the observed baryon asymmetry of the universe. Alternatively, it is possible to 
envisage generating a net lepton number $L$ at lower energies and subsequently producing the
baryon asymmetry via $B-L$ conservation~\cite{sphalerons, Fukugita:1986hr, Harvey:1990qw}.

In this paper we reconsider direct baryogenesis or baryogenesis via leptogenesis from the evaporation of primordial black holes.
Under the assumption that black holes dominate the energy density of the universe by the time they evaporate,
we present a simple derivation of the resultant asymmetry that is insensitive to the cosmological model. Hence, we can apply our 
results both to inflation~\cite{inflation} and ekpyrotic/cyclic models~\cite{justin, cyclic}. We find that the baryon asymmetry 
is independent of the initial black hole number density,
but does depend sensitively on the nature of the CP and $B$ or $L$-violating interactions.
We also consider the possibility that black hole evaporation stops due to quantum gravity effects when the mass reaches the Planck 
scale and the black holes form massive relics that act as dark matter.
The cosmological consequences of Planck-mass black hole relics have 
been studied previously by MacGibbon~\cite{MacGibbon}, Carr {\it et al.}~\cite{Carr:1994ar} and Barrow {\it et al.}~\cite{Barrow:1992hq}. 
Here we discuss the combined constraints if black holes form relics and also account for baryogenesis.

What processes in the early universe could have generated primordial black holes?
Primordial black hole formation from big bang inhomogeneities was first discussed by Carr and 
Hawking~\cite{Carr:1974nx}. 
However, if inflation removes all pre-existing inhomogeneities, any cosmologically interesting black hole 
density has to be created after inflation. A number of mechanisms have been proposed for generating post-inflationary primordial 
black holes.
Carr~\cite{blue} discussed black hole formation if the spectrum of density fluctuations generated during inflation is very blue 
(spectral index $n_s>1$).  If the primordial power spectrum is nearly-scale invariant and doesn't have significant running, then 
this mechanism is now tightly constrained by CMB measurements \cite{Spergel:2003cb}.
Primordial black hole production has also been proposed in the context of hybrid 
inflation. Garcia-Bellido {\it et al.}~\cite{Garcia-Bellido:1996qt} showed that density perturbations in hybrid inflation models can 
reach a large enough magnitude to produce black holes for wave numbers of order the horizon scale during 
the transition between the two inflationary stages. The mass and number 
density of black holes produced in these models is highly model-dependent, 
including large regimes which are cosmologically disastrous or insignificant. 
Hsu~\cite{Hsu} studied the formation of black holes in models of extended or hyperextended inflation 
(see \cite{extended}).  
These models contain two sources of fluctuations -- the usual nearly scale-invariant scalar field fluctuations and also 
inhomogeneities created by the collision of bubbles \cite{Hawking:ga} produced during the last few $e$-foldings. 
Barrow {\it et al.}~\cite{barrow} provided an analytical treatment of GUT baryogenesis from these black holes.

In this paper we present a model-independent derivation of the baryon asymmetry generated by primordial black holes that can be applied
in both inflationary and ekpyrotic/cyclic models. In the cyclic scenario black holes may form 
when branes collide. Since the reheat temperature in ekpyrotic/cyclic models is generally modest, it is important to have a reliable 
baryogenesis mechanism operating at temperatures well below the GUT-scale. 
We show that black hole baryogenesis provides a mechanism that is effectively 
independent of 
the reheat temperature of the universe. The potential challenge is that the ekpyrotic/cyclic model parameters must also satisfy a number of 
other conditions in order to resolve the horizon, flatness and monopole problems and to obtain an acceptable spectrum of density 
perturbations~\cite{Khoury:2003rt}. However, here we show that black hole baryogenesis is compatible with all current constraints for a 
wide range of parameters.

In both inflationary and ekpyrotic/cyclic models, precise calculations of the initial black hole density are either difficult or highly 
dependent on model parameters.
One of the important features of the scenario we are considering is that the baryon and 
possible remnant densities are insensitive to the initial black hole density provided the black holes dominate the universe by the time 
they decay. Reliable predictions for the baryon asymmetry can 
therefore be made 
even in the absence of an exact calculation of primordial black hole production.\\

The outline of the paper is as follows. 
In section \ref{sec:baryogenesis}, we examine black hole baryogenesis. 
After considering Hawking emission from a single black hole, we compute 
the baryon number
generated by an ensemble of black holes. 
We explore the parameter 
space of CP and $B$-violating interactions that reproduce observations.
Section \ref{sec:cyclic} applies our results to black hole baryogenesis in the ekpyrotic/cyclic model. 
In section \ref{sec:relics} we briefly discuss the cosmological implications if Planck mass remnants form as the final 
stage of black hole evaporation. We summarize our conclusions in section \ref{sec:discussion}.

We use natural units, $\hbar=c=k_B\equiv 1$, throughout, and define $8\pi G = 1/m_{\rm Pl}^2
\equiv 1$. For numerical estimates we use the reduced Planck mass
$m_{\rm Pl} \approx 10^{18}$ GeV and Planck time $t_{\rm Pl} = 1/m_{\rm Pl} \approx 5\times 10^{-44}$ sec.

\section{Black Hole Baryogenesis}\label{sec:baryogenesis}

We begin by giving a brief review of baryogenesis during Hawking evaporation~\cite{Hawking:sw}  of a 
single black hole (BH). Then we consider an ensemble of black holes and 
compute the resulting baryon asymmetry.

The horizon temperature of an uncharged, non-rotating Schwarzschild black hole is
\begin{equation}\label{equ:GH}
T_{\rm BH} = \frac{1}{M_{\rm BH}}\, .
\end{equation}
Its rate of evaporation is given by~\cite{Peacock:ye},\footnote{Note that Barrow {\it et al.}~\cite{barrow} use an approximate equation for the mass loss of a black hole which underestimates the true rate by a factor of $10^5$. This leads to an overestimation of the final baryon asymmetry by a factor of 100.}
\begin{equation}\label{equ:hawk}
\frac{d{M}_{\rm BH}}{dt} \approx -\frac{\pi^2}{120} g_* A_{\rm BH}  T_{\rm BH}^4 = - \frac{\pi}{480} \frac{g_*}{M_{\rm BH}^2}\, ,
\end{equation}
where $A_{\rm BH} = 4 \pi R_{\rm BH}^2$ is the surface area of the black hole event horizon, $M_{\rm BH}(t)$ is the instantaneous black hole mass and $g_* \sim 100$ is the 
effective number of relativistic degrees of freedom of particle species radiated at temperature $T_{\rm BH}$. 
Integrating equation (\ref{equ:hawk}) gives 
\begin{equation}\label{equ:mass}
M_{\rm BH}(t) =M_0 \left(1-\frac{t}{\tau}\right)^{1/3}\, ,
\end{equation}
where the black hole lifetime is
\begin{equation}\label{equ:life}
\tau =  \frac{160}{\pi} \frac{1}{g_*} M_0^3\, ,
\end{equation}
for a black hole with initial mass $M_0$.
The Gibbons-Hawking temperature (\ref{equ:GH}) implies the differential mass decrease 
\begin{equation}
dM_{\rm BH} = - \frac{1}{T_{\rm BH}^2} dT_{\rm BH} \equiv -dE\, ,
\end{equation}
where $dE$ is the energy emitted by the black hole when its mass decreases from $M_{\rm BH}$ to $M_{\rm 
BH} + dM_{\rm BH}$. 
Since the radiated particles have mean energy $3 T_{\rm BH}$, the 
differential number of particles emitted is
\begin{equation}\label{equ:diffno}
dN = \frac{dE}{3T_{\rm BH}} = \frac{1}{3} \frac{1}{T_{\rm BH}^3}  dT_{\rm BH}\, .
\end{equation}
Integrating equation (\ref{equ:diffno}) beginning from the initial black hole temperature $T_0 \equiv T_{\rm BH}(M_0)=1/M_0$ we find 
the total number of particles emitted by a single black hole of initial mass $M_0$ 
\begin{equation}\label{equ:N}
N= \int_{T_0}^{\infty} dN(T)  =\frac{M_0^2}{6}\, .
\end{equation}
Baryogenesis can occur through the emission of a particle species $X$ with 
mass $M_X$, whose subsequent decay violates baryon number conservation.
If the initial black hole mass is small enough, $T_0=1/M_0 > M_X$, the 
fraction of 
$X$ particles emitted is given by the equipartition factor
\begin{equation}\label{equ:f}
f_X \sim \frac{g_X}{g_*}\, ,
\end{equation}
where $g_X$ denotes the number of degrees of freedom of particle $X$.
The total number of $B$-violating particles is
\begin{equation}\label{equ:NX}
N_X = f_X N \quad \quad \text{if} \quad M_0 < \frac{1}{M_X}\, .
\end{equation}
Significant $X$-emission requires $T_{\rm BH} \ge M_X$.
Very massive black holes are initially not hot enough to emit $X$ particles ($T_0 < M_X$),
and baryon production during the initial stages of evaporation is exponentially suppressed. 
In this case, the lower limit ($T_0$) in the integral in equation (\ref{equ:N})
should be replaced by $T_{\rm BH} = M_X$, and
equation (\ref{equ:NX}) becomes
\begin{equation}\label{equ:NX2}
N_X = f_X \int_{T_X =M_X}^{\infty} dN(T) = f_X
\left(\frac{1/M_X}{M_0}\right)^2
\frac{M_0^2}{6} \quad \quad \text{if} \quad M_0 > \frac{1}{M_X}\, .
\end{equation}
Hence, we have
\begin{equation}\label{equ:Nx}
N_X = \kappa f_X N\, ,
\end{equation}
where
\begin{equation}\label{equ:kappa}
\kappa \equiv \left\{ \begin{array}{ll}
1 & \mbox{if $M_0 < 1/M_X$}\\
(M_0 M_X)^{-2 } & \mbox{if $M_0 > 1/M_X$} \end{array} \right.\, .
\end{equation}
$B$-violating $X$-decays may still lead to a baryon symmetric universe if the corresponding anti-particle processes occur at the same rate, {\it i.e.} baryon asymmetry requires CP violation.
If the CP violating parameter is
\begin{equation}\label{equ:gamma}
\gamma \equiv \sum_i B_i \frac{\Gamma (X\to f_i) - \Gamma(\bar{X} \to 
\bar{f}_i)}{\Gamma_{\rm tot}}\, ,
\end{equation}
where $B_i$ is the baryon number of the final state $f_i$, then
the excess
baryon number generated by the evaporation of a single black hole is 
\begin{equation}\label{equ:b}
\gamma N_X = \gamma \kappa f_X N\, .
\end{equation}

Next, we consider the baryon asymmetry generated by an ensemble of small black holes.
Imagine a dilute gas of primordial black holes, formed through some mechanism that we leave unspecified for the purpose of this 
general discussion. Depending on the details of the cosmological model, the universe may also contain a 
background of radiation and other contributions to the total energy density.  
As we shall see, the resulting baryon asymmetry is independent of these details provided  that 
the black holes survive long enough to dominate the universe by the time they evaporate.

In some of the cosmological models of interest,
including the extended inflation and ekpyrotic/cyclic models,
black holes are produced with a narrow distribution of masses set by the horizon size at the time of
formation.  Consequently,
we assume for simplicity an initial distribution in which all black holes have the
same mass $M_0$.  (It is straightforward to generalize our results to a broader distribution.)
Equation (\ref{equ:hawk}) shows that the most energy is transfered as the mass of the black hole becomes small during the final stages of the evaporation process. 
This and the fact that $t_{\rm formation} \ll \tau$ (where $\tau$ is the mean black hole lifetime) allows us to treat the evaporation as if 
all particles were produced at a single instant, $t_{\rm evap} \approx \tau$. 
Approximating equation (\ref{equ:mass}) by a step-function, $M_{\rm BH}(t) \approx M_0 \Theta(\tau -t)$, and assuming that 
black holes dominate the universe at the time they decay, we find
\begin{equation}\label{equ:massapprox}
\rho_{\rm evap} = \rho_{\rm BH}(t_{\rm evap}^{-}) \approx n_{\rm BH}^{\rm evap} M_0\, ,
\end{equation}
where $n_{\rm BH}^{\rm evap} \equiv n_{\rm BH}(t_{\rm evap})$ is the black hole number density at $t_{\rm evap}$. 
If the black holes dominate at evaporation then
the total entropy and baryon density of the universe is determined by the black hole evaporation products.
All black hole  energy density is transformed into radiation at $t_{\rm evap}$.
Hence, we have
\begin{equation}\label{equ:evap}
\rho_{\rm BH}(t_{\rm evap}^{-}) = \rho_{\gamma}(t_{\rm evap}^{+})\, .
\end{equation}
The radiated particles equilibrate with the surroundings and the temperature of the universe is defined via 
$\rho_{\gamma}^{\rm evap} \sim T_{\rm
evap}^4$. Using equations (\ref{equ:massapprox}) and (\ref{equ:evap}) we have
\begin{equation}\label{equ:energy}
n_{\rm BH}^{\rm evap} M_0 = \frac{\pi^2}{30} g_* T_{\rm
evap}^4\, .
\end{equation}
The Hubble parameter at evaporation (black hole domination) is given approximately by
\begin{equation}\label{equ:Hevap}
H_{\rm evap} \approx \frac{2}{3} \frac{1}{t_{\rm evap}}\, ,
\end{equation}
where $t_{\rm evap}$ is the FRW time. The last approximation is valid provided that $t_{\rm evap} \gg t_{\rm BH-domination}$.  
Using the black hole lifetime (\ref{equ:life}) and $H_{\rm evap}^2 = \rho_{\rm evap}/3$ in combination with equation (\ref{equ:massapprox}) 
gives
\begin{equation}\label{equ:hubble}
H_{\rm evap}^2 = \frac{4}{9} \left(\frac{\pi}{160}\right)^2 \frac{g_*^2}{M_0^6} = \frac{1}{3} n_{\rm BH}^{\rm evap} M_0\, .
\end{equation} 
Solving equation (\ref{equ:hubble}) for the black hole  number density at evaporation we get
\begin{equation}\label{equ:nevap}
n_{\rm BH}^{\rm evap} = \frac{25}{48} \pi^2 \cdot \left(\frac{g_*}{100}\right)^2 \cdot \frac{1}{M_0^7}\, .
\end{equation} 
Note that the black hole number density at evaporation is independent of both the number density at formation and the cosmological model. 
This is a direct consequence of the assumption that the black holes come to dominate the universe at some point in the evolution.
It means that the cosmological conditions at evaporation that determine both the entropy and the baryon number are characterized by a 
single parameter, $M_0$.

From equation (\ref{equ:energy}) we then obtain the evaporation temperature
\begin{equation}\label{equ:Tevap}
T_{\rm evap} = \left(\frac{5}{32}\right)^{1/4} \cdot \left(\frac{g_*}{100}\right)^{1/4} \cdot \frac{1}{M_0^{3/2}}\, .
\end{equation}
In order to reproduce the success of standard big-bang cosmology, we require the universe to be radiation dominated at 
nucleosynthesis. The black holes therefore must have evaporated by that time, which leads to the following {\it BBN constraint}
\begin{equation}
T_{\rm evap} > T_{\rm BBN} \sim 10^{-22}\, .
\end{equation}
As a limit on the initial black hole  mass, this gives
\begin{equation}
M_0 < 4 \times 10^{14} \left( \frac{g_*}{100}\right)^{1/6}\, .
\end{equation}
Using equation (\ref{equ:b}) for the net baryon number created by a single black hole, the number 
density of baryons at the time of evaporation is
\begin{equation}\label{equ:nB}
n_B \equiv n_B(t_{\rm evap})= \gamma \kappa f_X N n_{\rm BH}^{\rm evap}\, .
\end{equation}
We define the effective entropy radiated by a single black hole as
\beq
\sigma \equiv \frac{s}{n_{\rm BH}^{\rm evap}}\, ,
\eeq
where the entropy density generated by a gas of black holes is
\begin{equation}\label{equ:entr}
s \equiv s(t_{\rm evap}) = \frac{2 \pi^2}{45} g_{*} T_{\rm evap}^3\, .
\end{equation}
Combining (\ref{equ:nB}) and (\ref{equ:entr}), the baryon-to-entropy ratio is
\begin{equation}\label{equ:baryons}
B = \frac{n_B}{s} = \frac{\gamma \kappa f_X N}{\sigma}\, .
\end{equation}
Equations (\ref{equ:entr}), (\ref{equ:nevap}) and (\ref{equ:Tevap}) give the entropy produced per black hole 
\begin{equation}\label{equ:sigma2}
\sigma = \frac{4}{3} \frac{M_0}{T_{\rm evap}} = \frac{4}{3} \left(\frac{32}{5}\right)^{1/4} \left(\frac{g_*}{100}\right)^{-1/4} \cdot 
M_0^{5/2} \, .  
\end{equation}
This shows that the effective entropy radiated by a single black hole is a factor of $\sqrt{M_0}$ larger than the initial 
Bekenstein-Hawking 
entropy of the black hole, $S=\sqrt{\frac{\pi}{2}} A_{\rm BH} \sim M_0^2$. 
Substituting $N=M_0^2/6$ and $\sigma(M_0)$ into (\ref{equ:baryons}), we finally get our master expression for the baryon 
number of the universe
\begin{equation}\label{equ:finalA}
B \approx \left[ 0.1 \left(\frac{g_*}{100}\right)^{1/4} f_X\right] \cdot \frac{\gamma}{M_0^{1/2}} \kappa
\, .
\end{equation} 
The parameters $g_*$, $\gamma$, $f_X$, and $M_X$ are set by microphysics, whereas $M_0$ is determined by the cosmological model.
Using $f_X \sim 1/g_*$ the first factor in (\ref{equ:finalA}) is $C(g_*) \equiv 10^{-3} 
\left(\frac{g_*}{100}\right)^{-3/4}$.
Note that $B$ is independent of the initial black hole  number density. This illustrates that the final baryon asymmetry is 
insensitive to the initial conditions as long as evaporation occurs after the black holes dominate. 
The final baryon asymmetry (\ref{equ:finalA}) displays a simple dependence on the free parameters $M_0$, $\gamma$ and $M_X$. $B$ scales 
linearly with $\gamma$, the net baryon number created by each black hole. For small initial black hole  
masses, $M_0 < 1/M_X$, $\kappa=1$ and $B$ is simply proportional to 
the ratio $\gamma/M_0^{1/2}$. Increasing $M_0$ therefore suppresses baryon production. Very large initial masses, $M_0 > 1/M_X$, lead to 
a further suppression from the factor $\kappa = 1/(M_0 M_X)^2 < 1$.\\

\begin{figure}[h!]
  \centering
   \includegraphics[width=0.65\textwidth]{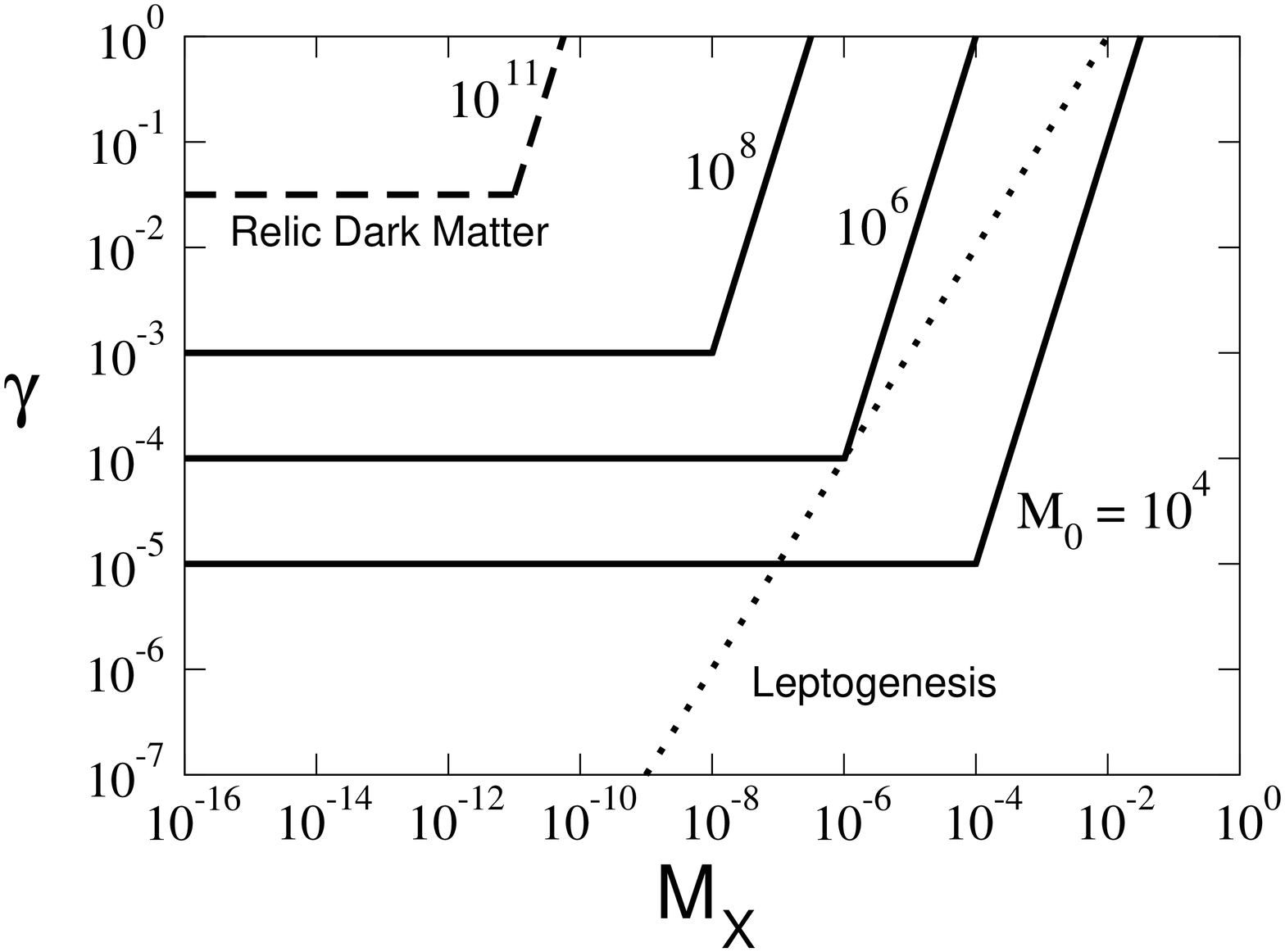} 
\caption{[$B$=$10^{-10}$]--contours in the $\gamma$--$M_X$ plane for three different values of the initial black hole mass $M_0$ ({\it cf.} 
equation (\ref{equ:finalA})). $\gamma$ determines the amount of CP violation, while $M_X$ is the mass of the $B$-violating particle $X$.
The contours correspond to parameter combinations that produce $B \approx 10^{-10}$.
Regions 
below the contours lead to a baryon number that is too small to be consistent with observations ($B<10^{-10}$). Regions above the 
contours overproduce baryons ($B > 10^{-10}$). Since entropy production at a later stage of evolution may dilute $B$, this region is viewed 
as a cosmologically acceptable region. The dotted 
line corresponds to the leptogenesis model of \cite{Hamaguchi:2001gw} which predicts $\gamma = 10^2 M_X$. The dashed line defines the region of parameter space which, 
in principle, allows black holes to simultaneously be the source of baryons and dark matter (see section \ref{sec:relics}). Note that this requires 
very large CP violation for small $M_X$.}
\label{fig:gamma}
\end{figure}

Figure \ref{fig:gamma} is a plot of [$B$=$10^{-10}$]--contours in the $\gamma$--$M_X$ plane for three different fixed values of the 
initial black hole mass $M_0$. 
For $M_X < 1/M_0$ (immediate emission of $X$ particles) $\gamma = \frac{B}{C} M_0^{1/2} = \text{const}$, 
independent of $M_X$, giving a horizontal section. For more massive $X$ particles, $M_X > 1/M_0$, baryogenesis is suppressed and 
$\gamma$ has to be increased to compensate that suppression. 
This 
explains the rise of the contours for large $M_X$. The transition between these two regimes occurs at $M_X = 1/M_0$. Each contour 
corresponds to combinations of the three parameters $\gamma$, $M_X$ and 
$M_0$ that reproduce the observed baryon asymmetry, $B \approx 10^{-10}$. 
The region below the contours corresponds to parameter combinations that are unacceptable because they underproduce baryons, $B<10^{-10}$. This is the 
forbidden region. The region above the contours overproduces baryons, $B>10^{-10}$. We
call this the allowed region. 
Overproduction is allowed since we can imagine  non-adiabatic processes after the black holes evaporate that
create extra entropy and dilute the abundance to the correct value. We see that there is a wide range of allowed parameters.
We also notice that increasing $M_0$ decreases the allowed region. Small masses, therefore, seem to be preferable, although consistency with the 
assumption that the black holes live long enough to dominate the universe at evaporation has to be tested separately. This can only be 
done once the cosmological setup is specified.

Thus far, we treated $\gamma$ and $M_X$ as independent free parameters. We now show how this plot can be used to explore a given 
particle physics model of baryogenesis that relates $\gamma$ and $M_X$. For example, we consider the 
leptogenesis model in \cite{Hamaguchi:2001gw} that relates the CP violation parameter $\gamma$ to $M_X$:
\begin{equation}\label{equ:lepto}
\gamma = 10^2 M_X\, .
\end{equation}
Using the above analysis we can easily evaluate the feasibility of this black hole leptogenesis scenario. 
The dotted line in Figure~\ref{fig:gamma} corresponds to equation (\ref{equ:lepto}).
For small $M_0$ there is a finite range of $\gamma(M_X)$ in the region that allows sufficient baryogenesis. In Figure 
\ref{fig:gamma} this is illustrated for $M_0=10^4$. The dotted line crosses the [$B=10^{-10}$]--contour at two points. These points 
correspond to combinations of $\gamma(M_X)$ and $M_0$ values that reproduce observations precisely. In between these two points there is a finite line segment 
where baryons are 
overproduced, which is also allowed assuming some modest dilution later. 
For this particular black hole leptogenesis model to be feasible, we conclude that
\begin{equation}
M_0 < 10^6\, ;
\end{equation}
otherwise, the dotted line will lie completely in the forbidden region and baryons are underproduced for all choices of parameters.

\section{Black Hole Baryogenesis in the \\Ekpyrotic/Cyclic Models}\label{sec:cyclic}

Our derivation of black hole baryogenesis is insensitive to the cosmological model provided that the black holes dominate the energy density of the universe at the time of evaporation. An interesting application is to the recently 
proposed ekpyrotic/cyclic models~\cite{justin, cyclic}.
According to the ekpyrotic model the hot big bang is caused by the collision of two 3-branes bounding an extra spatial 
dimension~\cite{justin}. In the cyclic scenario this may repeat itself periodically, leading to a periodic sequence of big bangs and 
big crunches. The model is designed to resolve 
the cosmological horizon, flatness and monopole problems and generate a nearly scale-invariant spectrum of density perturbations 
without invoking a period of high-energy inflation~\cite{inflation}. 
The ekpyrotic/cyclic model may also provide a natural setting for black hole baryogenesis.
The ekpyrotic mechanism is designed to smooth the universe on superhorizon scales before the
bounce occurs.  At present, a bounce is not proven to occur, but, assuming it does, it is reasonable
to suppose that instabilities develop within the horizon and produce black holes on subhorizon scales. In this section we consider whether black hole baryogenesis is compatible with the constraints on
ekpyrotic/cyclic models that are already known~\cite{Khoury:2003rt}.\\

\begin{figure}[h!]
  \centering
   \includegraphics[width=0.90\textwidth]{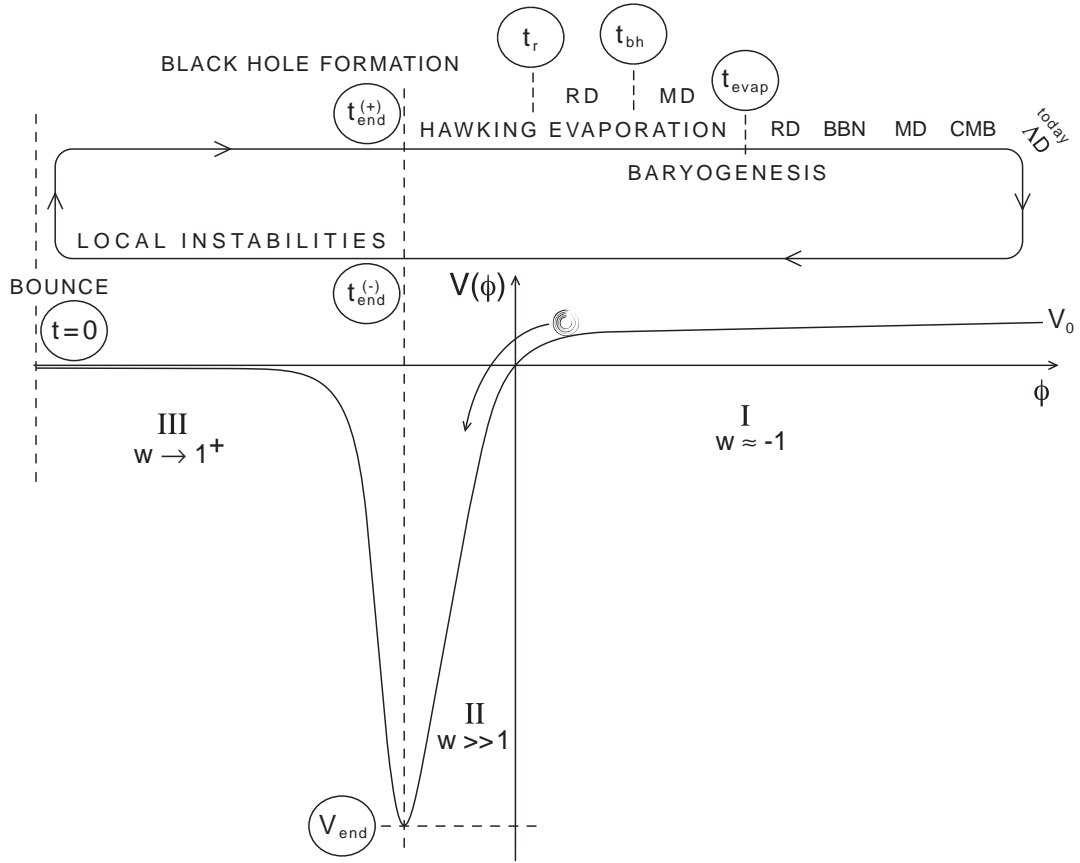} 
\caption{Black hole baryogenesis in the ekpyrotic/cyclic universe. The scalar field $\phi$ parameterizes the distance between the branes. 
$V(\phi)$ is a generic potential for $\phi$. Region I corresponds to the observed dark energy domination today. A scale-invariant spectrum of 
perturbations is generated in region II. Perturbations are stabilized until $t_{\rm end}^{(-)}$, after which local instabilities develop 
and form black holes. The size of these black holes is limited by the horizon size at $t_{\rm end}^{(-)}$. The cosmological implications 
of these black hole for the post-big bang evolution of the universe are studied in this paper.} \label{fig:pot} 
\end{figure}

According to the cyclic model, the universe is dominated by the brane kinetic energy immediately after the bounce. 
In the effective 4D theory, the separation between the branes is described by a scalar field $\phi$ and the brane kinetic energy is 
characterized by the scalar field kinetic energy $\frac{1}{2} \dot{\phi}^2 \propto a(t)^{-6}$. 
A generic potential for $\phi$ is shown in Figure~\ref{fig:pot}.
Region I corresponds to the present epoch, which is characterized by dark energy domination and accelerated expansion. 
Scale-invariant density
perturbations are created during a phase of slow contraction (region II). Chaotic behavior in the big crunch is suppressed by the
exponential form of the potential ($w \gg 1$, see \cite{joel}) and fluctuations are linear until $t_{\rm end}^{(-)}$. However,
short-wavelength fluctuations become highly non-linear after $t_{\rm end}^{(-)}$ (region III). The distance between the branes becomes 
small and 
quantum gravity corrections become
important. Classical general relativity is insufficient to describe the further evolution. It is known, however,  that the effective 
gravitational constant becomes large when the branes collide and particle collisions
reach large energies. 
The horizon at time $t$ sets the maximal distance a signal sent at time $t$
can travel prior to the bounce.
We therefore assume that the instabilities that
develop in the big crunch are limited by causality to small scales, the maximum scale being the horizon at $t_{\rm 
end}^{(-)}$ (the time when perturbative control is lost). This sets the maximum black hole mass, but we don't call it a 
black hole before the bounce, since the cosmological horizon shrinks to become smaller than the Schwarzschild horizon. 
The horizon grows after the bounce to encompass the instability at about $t_{\rm end}^{(+)}$ (see Figure~\ref{fig:pot}). We 
define black hole formation as the time $t_f$ after the bounce when $H^{-1}(t_f) = H_{\rm end}^{-1} \equiv 
H^{-1}(t_{\rm end}^{(-)})$, {\it i.e.}, the horizon becomes comparable to the Schwarzschild radius. The near symmetry of the 
time evolution of the scalar field ($H(t_{\rm end}^{(-)}) \approx H(t_{\rm end}^{(+)})$) then allows us to use the 
horizon at $t_{\rm end}^{(-)}$ as a scale for the black 
hole formed at $t_f \approx t_{\rm end}^{(+)}$.

\begin{figure}[h!]
  \centering
   \includegraphics[width=0.75\textwidth]{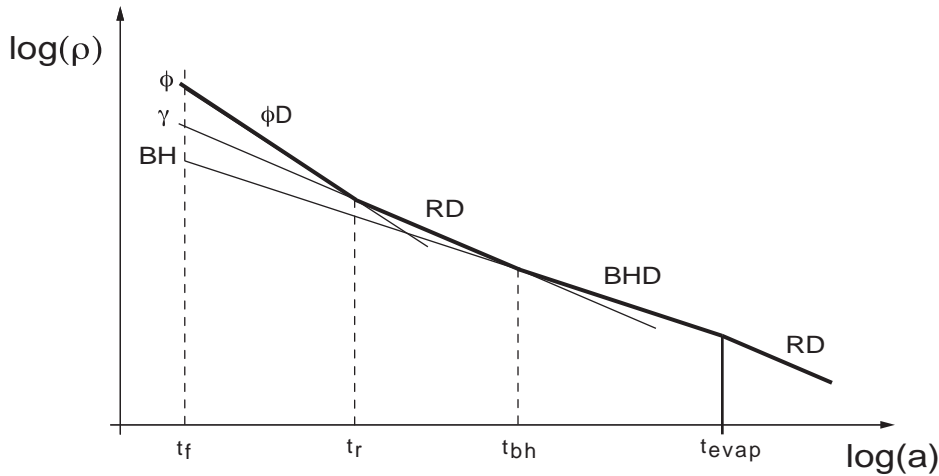} 
\caption{
In the cyclic model, the universe changes from scalar field dominated ($\phi$D) when 
the black holes form (at $t=t_f \approx t_{\rm end}^{(+)}$), to radiation dominated (RD) at $t_r$, to black hole 
dominated (BHD) at $t_{\rm bh}$, 
and back 
to radiation dominated at $t_{\rm evap}$.}
\label{fig:plot}
\end{figure}

Here we focus on the evolution after the bounce.
Assuming the collision produces significant amounts of radiation and black holes, the various densities evolve according to
Figure \ref{fig:plot}.
The case we study in detail below (and that is illustrated in Fig.~\ref{fig:plot}) assumes for simplicity that the initial radiation 
density produced at the collision overtakes the scalar field kinetic energy before the
black holes. 
This is equivalent to assuming that the energy
density in black holes at radiation--scalar field equality ($t_r$) is less than the energy density in radiation. We later reformulate
this as a constraint on the number of black holes  per horizon at the time of formation. Since radiation dilutes more quickly than the
pressureless gas of black holes, the black holes will eventually come to dominate the universe ($t > t_{\rm bh}$).
As we have shown, the baryon asymmetry resulting from their evaporation is then independent of the initial black hole number 
density, but depends sensitively on the nature of the CP and $B$ or $L$-violating interactions.
The initial radiation energy density redshifts away in the subsequent black hole dominated phase and plays no significant
role afterwards. Black holes dominate for a finite time until they decay at $t_{\rm evap}$.
Evaporation occurs before nucleosynthesis ($t_{\rm evap} < t_{\rm BBN}$), so standard cosmology is recovered after that time.

The previous discussion suggests that the bounce produces both a small amount of radiation from the non-adiabatic collision of the
branes and a gas of primordial black holes.
Once the radiation or black hole dominated epochs begin, the scalar field kinetic energy is rapidly damped and the field comes to halt. 
In order to cycle, radiation or black hole domination should not begin until $\phi$ has had time to go from 
$- \infty$, past $V=-\left|V_{\rm end}\right|$, across the potential well and back up the potential to the plateau where $V={\cal 
O}(V_0)$ (see Fig.~\ref{fig:pot}). Otherwise the scalar field would get trapped in the minimum at $V_{\rm end}$ and the universe would be 
anti-de Sitter space.
Khoury {\it et al.}~\cite{Khoury:2003rt} showed that this leads to the following {\it cycling constraint}
on the radiation energy density (or temperature) at scalar kinetic energy-radiation equality ($t_r$)
\begin{equation}\label{equ:cyc}
\rho_r \sim T_r^4 \le V_{\rm end} \left(\frac{V_0}{V_{\rm end}}\right)^{\sqrt{6\varepsilon}}\, ,
\end{equation}
where $\varepsilon$ is the fast-roll parameter defined in ekpyrotic cosmology~\cite{cyclic}.
Under the assumption that radiation from the collision dominates before the black holes, we have
\begin{equation}
\rho_{\rm BH}^r < \rho_{\gamma}^r \sim \rho_r\, .
\end{equation}
This can be re-expressed as a condition at the time of black hole formation.
The black hole  number density at radiation--scalar field equality 
is related to the density at formation via
\begin{equation}
n_{\rm BH}^r = n_{\rm BH}^f \left(\frac{a_f}{a_r}\right)^3 = n_{\rm BH}^f \frac{H_r}{H_f}\, ,
\end{equation}
where we used $H^2 \propto a^{-6}$ during domination of scalar field kinetic
energy ($t_f \approx t_{\rm end}^{(+)} \to t_r$). 
Assuming that the initial black hole  mass is equal to the horizon 
mass at formation, $M_0 \sim \rho_f H_f^{-3} \sim 
H_f^{-1}$, and using $H_r^2 \sim T_r^4$ gives
\begin{equation}
n_{\rm BH}^r = n_{\rm BH}^f M_0 T_r ^2\, .
\end{equation}
Defining the number of black holes per horizon at formation as
\begin{equation}
N_{\rm BH} \equiv n_{\rm BH}^f M_0^3\, ,
\end{equation}
we find
\begin{equation}\label{equ:nBHr}
n_{\rm BH}^r = N_{\rm BH} \cdot \frac{T_r^2}{M_0^2}\, .
\end{equation}
Imposing the cycling constraint
in the form
$\rho_{\rm BH}^r \sim n_{\rm BH}^r M_0 < \rho_{\gamma}^r \sim T_r^4$,
we obtain
\begin{equation}\label{equ:cycling}
N_{\rm BH} < M_0 T_r^2\, ,
\end{equation}
where $T_r$ obeys the cycling condition (\ref{equ:cyc}). Black hole baryogenesis works for larger $N_{\rm BH}$, but then black holes 
dominate before the collisional radiation and a separate analysis is needed.

We also impose the constraint that the black holes survive long enough to eventually dominate the universe. This is required for the results of section \ref{sec:baryogenesis} to be applicable.
We denote quantities evaluated at radiation--black hole equality by (bh). The condition that the black holes survive long enough to dominate the universe 
is $\tau > t_{\rm bh}$,
or in terms of the scale factor of the universe
\begin{equation}
a_{\rm evap} > a_{\rm bh}\, .
\end{equation}
At radiation--black hole equality we have,
$\rho_{\rm BH}(t_{\rm bh}) \equiv \rho_{\gamma}(t_{\rm bh})$ ,
or
\begin{equation}\label{equ:radBH}
n_{\rm BH}^r M_0 \left( \frac{a_r}{a_{\rm bh}}\right)^3 \equiv T_r^4 \left( 
\frac{a_r}{a_{\rm bh}}\right)^4\, .
\end{equation}
From (\ref{equ:radBH}) and (\ref{equ:nBHr}) we obtain
\begin{equation}
\frac{a_r}{a_{\rm bh}} = \frac{N_{\rm BH}}{M_0 T_r^2}\, .
\end{equation}
This confirms that the cycling condition (\ref{equ:cycling}) is equivalent to $a_{\rm bh} > 
a_r$. Using $a_r/a_{\rm bh}=T_{\rm bh}/T_r$ we get the temperature of the universe at 
radiation--black hole equality
\begin{equation}
T_{\rm bh} =\frac{N_{\rm BH}}{M_0 T_r}\, .
\end{equation}
During black hole domination ($t_{\rm bh} \to t_{\rm evap}$) we have
\begin{equation}
\left( \frac{a_{\rm evap}}{a_{\rm bh}}\right)^3 = \frac{H_{\rm bh}^2}{H_{\rm evap}^2} = 
\frac{T_{\rm bh}^4}{1/\tau^2} = M_0^6 \frac{N_{\rm BH}^4}{M_0^4 
T_r^4}\, . 
\end{equation}
Notice that $H_{\rm evap}$ is related to the lifetime of the black holes.  
The condition that the black holes  survive long enough to dominate the universe, $a_{\rm evap} > a_{\rm bh}$, translates into $H_{\rm evap} < H_{\rm bh}$, or
\begin{equation}\label{equ:survival}
N_{\rm BH} > \frac{T_r}{M_0^{1/2}}\, .
\end{equation}
This gives a lower limit on the initial black hole  mass
\begin{equation}\label{equ:survival2}
M_0 > \frac{T_r^2}{N_{\rm BH}^2}\, ,
\end{equation}
where $T_r$ is bounded by the cycling condition.
$T_r$ is determined by the initial ratio of scalar kinetic energy and radiation density,
with larger $T_r$ corresponding to more radiation produced at the bounce.
Keeping everything else fixed, while increasing $T_r$, increases the time to radiation--black hole  equality. As (\ref{equ:survival2}) 
shows, the black holes then have to be more massive, in order to survive long enough to dominate. 
$N_{\rm BH}$ is
related to the number of black holes  at formation. For a large number of primordial black holes,  radiation-black hole equality occurs earlier.
Hence, increasing $N_{\rm BH}$ while keeping $T_r$ (or the initial amount of radiation) fixed, decreases the time to radiation--black hole  
equality. The constraint on $M_0$ therefore weakens.

Combining the cycling (\ref{equ:cycling}) and survival (\ref{equ:survival}) conditions gives
\begin{equation}
\frac{T_r}{M_0^{1/2}} < N_{\rm BH} < M_0 T_r^2\, ,
\end{equation}
where
\begin{equation}
T_r \le V_{\rm end}^{1/4} \left( \frac{V_0}{V_{\rm end}} \right)^{\sqrt{3 
\varepsilon/8}}\, .
\end{equation}
Hence, cycling and survival can simultaneously be achieved only if the following {\it consistency constraint} is satisfied
\begin{equation}\label{equ:cons}
M_0 > T_r^{-2/3}\, .
\end{equation}
As discussed earlier, the maximal initial black hole mass is set by the horizon mass at the end of the ekpyrotic phase:
\begin{equation}
M_0 \le M_{\rm Hor}(t_{\rm end}^{(-)})  \sim \rho(t_{\rm end}^{(-)}) \left|H_{\rm end}\right|^{-3} \sim \left|H_{\rm end}\right|^{-1}\, .
\end{equation}
In \cite{Khoury:2003rt} it was shown that the horizon at $t_{\rm end}$ is
\begin{equation}
\left|H_{\rm end}\right|^{-1} \sim (2 \varepsilon V_{\rm end})^{-1/2}\, .
\end{equation}
This gives the following relation between $M_0$, $\varepsilon$, and $V_{\rm end}$
\begin{equation}\label{equ:M}
M_0 \le (2 \varepsilon V_{\rm end})^{-1/2}\, .
\end{equation}
Hence, we have related the initial
mass of black holes to the depth of the ekpyrotic potential.\\

\begin{figure}[h!]
  \centering
   \includegraphics[width=0.65\textwidth]{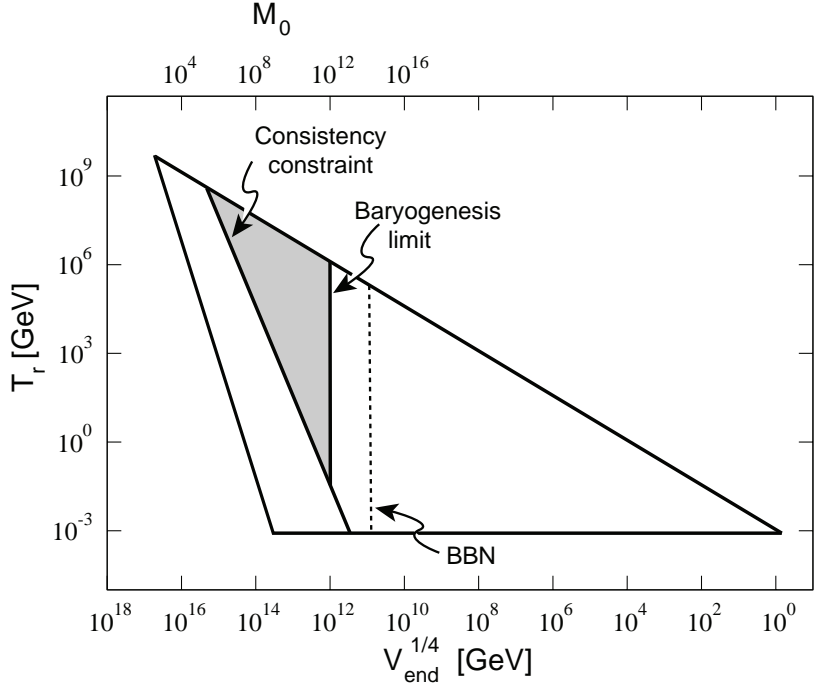} 
\caption{Constraints on the cyclic model with (grey triangle)
and without (outer triangle) black hole baryogenesis: 
The outer triangular region (reproduced from \cite{Khoury:2003rt}) shows the range of $V_{\rm end}^{1/4}$ and the reheat temperature 
$T_r$, 
over which the cyclic model satisfies all known cosmological constraints, 
with fixed $\varepsilon = 10^{-2}$.
Relating the initial black hole mass $M_0$ to the horizon mass at the end of the ekpyrotic phase leads to an estimate of 
$M_0$ in terms of the depth of the ekpyrotic potential $V_{\rm end}$: $M_0 \sim V_{\rm end}^{-1/2}$. The black hole 
baryogenesis analysis of section \ref{sec:baryogenesis} is only applicable if $M_0 > T_r^{-2/3}$, or $V_{\rm end}^{1/4} 
< T_r^{1/3}$. This is labeled the {\it consistency constraint}. Imposing an upper bound on the CP 
violation parameter $\gamma < 10^{-1}$ and using $\gamma_{\rm min} \approx 10^{-7} M_0^{1/2}$ from section 
\ref{sec:baryogenesis} gives the {\it baryogenesis limit}: $M_0 < 10^{12}$ or $V_{\rm end}^{1/4} > 10^{12}$ GeV. The 
grey region corresponds to parameter combinations for which primordial black hole baryogenesis is viable and consistent 
with known constraints on the cyclic model.}
\label{fig:design}
\end{figure}

Figure \ref{fig:design} (adapted from \cite{Khoury:2003rt}) illustrates the region of parameter space for which the cyclic model 
satisfies all known cosmological constraints. The outer triangular region in the $V_{\rm end}$--$T_r$ plane encloses the range of 
parameters that satisfy the BBN, cycling and gravitational wave background constraints studied in \cite{Khoury:2003rt}. 
For the purpose of illustration, assuming that the black hole masses saturate the bound 
(\ref{equ:M}), {\it i.e.} $M_0 = (2 \varepsilon V_{\rm end})^{-1/2}$, 
 we consider the constraint set by black hole baryogenesis. We impose an upper bound on CP violation, $\gamma < 10^{-1}$. 
From the analysis presented in section \ref{sec:baryogenesis} ($\gamma_{\rm min} \approx  10^{-7} M_0^{1/2}$) this implies 
$M_0< 10^{12}$. Using the relation between $M_0$ 
and $V_{\rm end}$ we get the baryogenesis limit
\begin{equation}\label{equ:limit1}
V_{\rm end}^{1/4} > \left(2 \varepsilon\right)^{-1/4} 10^{-6} \approx 10^{-6} = 10^{12} \, \, \text{GeV}\, ,
\end{equation}
for $\varepsilon \sim 10^{-2}$. In addition, we can re-express the consistency condition (\ref{equ:cons})
\begin{equation}\label{equ:limit2}
V_{\rm end}^{1/4} < \left(2 \varepsilon\right)^{-1/4} T_r^{1/3}\, .
\end{equation}
Equations (\ref{equ:limit1}) and (\ref{equ:limit2}) together with the constraints from \cite{Khoury:2003rt} define the grey region in Figure \ref{fig:design}. This corresponds to the region of parameter space for which the black hole baryogenesis mechanism of section {\ref{sec:baryogenesis}} is viable. We point out that regions outside the grey area are not strictly disallowed, but correspond to regions of parameter space that either require a separate analysis or where insufficient information is available to assess the viability of black hole baryogenesis.
The consistency constraint is not a real physical constraint, but merely the limit of 
applicability of our 
analysis. In that sense the shaded area may be viewed as a minimal allowed region. Finally, the allowed region should be interpreted in combination with the generic results of section 
\ref{sec:baryogenesis} (Figure \ref{fig:gamma}). These imply that large $M_0$ (small $V_{\rm end}$) may require unconventional particle physics 
to give the large value of $\gamma$ required to obtain $B=10^{-10}$. Successful black hole baryogenesis is therefore more physically plausible in this sense towards the left (increasing $V_{\rm 
end}$/decreasing $M_0$) of the allowed region. 

\section{Black Hole Remnants as Dark Matter?}\label{sec:relics}

Hawking's semi-classical treatment of black hole evaporation breaks down 
as the mass of the black hole approaches the Planck mass and quantum gravitational effects become large. 
A complete theory of quantum gravity seems necessary to 
decide on the final state of black hole evaporation. So far we have assumed that the
black holes evaporate completely and the final black hole temperature tends 
to infinity. We now extend the discussion to include the possibility that the black holes cease evaporating when the mass is of the 
order the Planck scale due to quantum gravity effects and form stable Planck mass relics. 
Whether this concept is plausible or not is beyond the scope of this paper (but see \cite{relics}). Here we simply consider whether such 
remnants, if they form, could account for the dark matter.\\

If $\Omega_i$ is the ratio of the density of component $i$ to the critical density, then dark matter--radiation equality occurs when
$\frac{\Omega_{\rm DM}}{\Omega_{\gamma}} = \frac{\Omega_{\rm DM,0}}{\Omega_{\gamma,0}} 
\, \frac{a}{a_0} \equiv 1$,
and the temperature at equality is
\begin{equation}\label{equ:TDM}
T_{{\rm DM=}\gamma} = \frac{\Omega_{{\rm DM},0}}{\Omega_{\gamma,0}} T_0\, ,
\end{equation}
where $T_0$ is the CMB temperature today.
From black hole evaporation, the densities of radiation and relics are 
\begin{eqnarray}
\rho_{\rm rel}^{\rm evap} &=& n_{\rm BH}^{\rm evap} M_{\rm rel}\, , \label{equ:Nrel}\\
\rho_{\gamma}^{\rm evap} &=& n_{\rm BH}^{\rm evap} M_0\, . \label{equ:Nrel2}
\end{eqnarray}
Note that in (\ref{equ:Nrel}) we assume for simplicity that the initial mass spectrum has a very narrow range about $M_0$ and can therefore be approximated by a $\delta$-function. \\ 
Equations (\ref{equ:Nrel}) and (\ref{equ:Nrel2}) imply relic--radiation equality at 
$\frac{ \rho_{\rm rel}^{\rm evap} }{ \rho_{\gamma}^{\rm evap} } \frac{a_{{\rm rel=}\gamma}}{a_{\rm 
evap}} \equiv 1$,
with corresponding temperature 
\begin{equation}\label{equ:Trel} 
T_{{\rm rel=}\gamma} = \frac{M_{\rm rel}}{M_0} T_{\rm evap}\, .
\end{equation}
If the black hole relics are the dark matter, then 
\begin{equation}
T_{{\rm rel=}\gamma} = T_{{\rm DM=}\gamma}\, ,
\end{equation}
or, from equations (\ref{equ:TDM}) and (\ref{equ:Trel}),
\begin{equation}\label{equ:2}
\frac{ \Omega_{{\rm DM},0} }{\Omega_{\gamma,0}} = \frac{M_{\rm rel}}{M_0} \frac{T_{\rm evap}}{T_0}\, .
\end{equation}
Substituting $T_{\rm evap}$ from equation (\ref{equ:Tevap}) gives
\begin{equation}
\frac{\Omega_{{\rm DM},0}}{\Omega_{\gamma,0}} =  \frac{M_{\rm rel}}{T_0} \left(\frac{5}{32}\right)^{1/4} \left(\frac{g_*}{100}\right)^{1/4} \frac{1}{M_0^{5/2}}\, .
\end{equation}
Assuming $M_{\rm rel} \sim 1$, $g_* \sim 100$, and the WMAP results \cite{Spergel:2003cb},
$\frac{\Omega_{{\rm DM},0}}{\Omega_{\gamma,0}} \approx 10^4$ and $T_0 = 2.7 \, \text{K} \sim 10^{-31}$,
we obtain
\begin{equation}
\left(M_0\right)_{\rm DM} \sim 10^{11}\, .
\end{equation}
Black hole relics can be the dark matter if the characteristic initial black hole mass has this virtually unique value.
If $M_0 < 10^{11}$ then $\Omega_{\rm rel} > \Omega_{\rm DM}$, and relics are overproduced compared to what
observations allow. Hence, if stable black hole relics are the final stage of Hawking evaporation then their initial mass 
must 
be greater than 
$(M_0)_{\rm DM} 
= 10^{11}$, an important constraint on black hole baryogenesis.
 
 \begin{figure}[h!]
   \centering
   \includegraphics[width=0.65\textwidth]{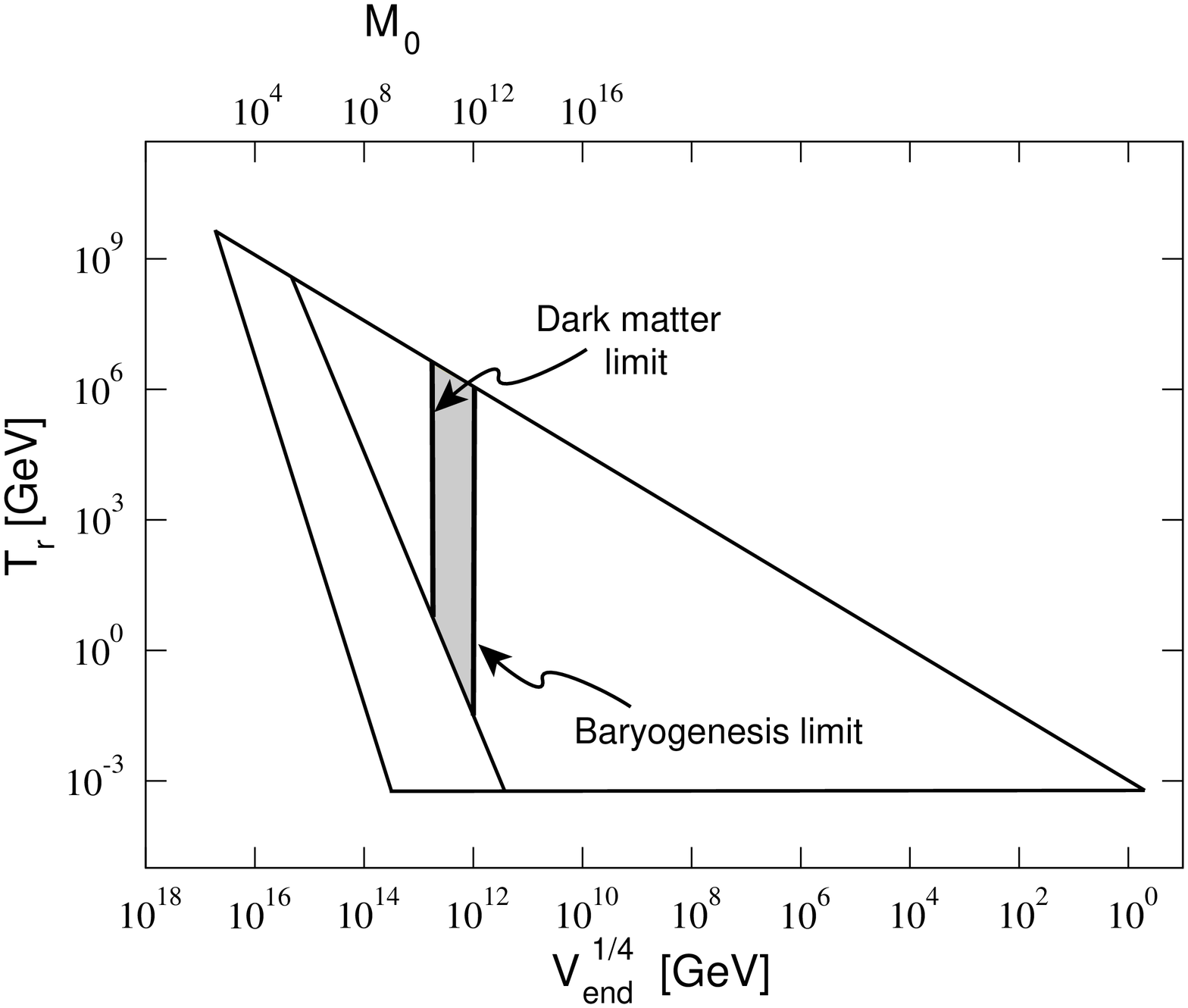} 
\caption{Constraints on the ekpyrotic/cyclic model: Same as Figure \ref{fig:design}, but with relic dark matter limit included. The grey region is the new 
allowed region.}
\label{fig:design2}
\end{figure}

It is interesting to consider whether the black holes could simultaneously be the source of dark matter and
baryon asymmetry. 
For a generic black hole baryogenesis scenario this is easily analyzed by reconsidering Figure \ref{fig:gamma}.
The dashed line in Fig. \ref{fig:gamma} is the limit on $\gamma$--$M_X$ for $M_0 =10^{11}$. 
The region above the dashed line in principle allows primordial black holes to be the source of both the baryon asymmetry and the 
dark matter of the universe. Note, however, that this requires a CP violation parameter of $\gamma > 10^{-2}$ for $M_X < 1/M_0 
= 10^{-11}$. 
Typical particle physics models of CP violation do not predict such large values for $\gamma$, which
 suggests that black hole baryogenesis and relic dark matter can't be realized simultaneously.\\
Incorporating the analysis of black hole relics into our constraints on black hole baryogenesis in the cyclic 
model leads to Figure \ref{fig:design2}. Only the parameter space corresponding to large black hole masses and large $\gamma$ is 
now allowed. We emphasize that it seems very hard to conceive a realistic particle physics model of CP violation in this regime.

\section{Discussion}
\label{sec:discussion}

Hawking evaporation of primordial black holes provides an interesting mechanism for generating the baryon asymmetry of the universe.
In this paper we presented a discussion of primordial black hole baryogenesis that is insensitive to the background cosmology and investigated the 
parameter space for which the theory reproduces observations.
A gas of small primordial black holes may be produced on subhorizon scales in the ekpyrotic/cyclic models. Applying our model-independent
results for black hole baryogenesis to the ekpyrotic/cyclic model we showed that,
for a large range of model parameters consistent with all other known constraints, these black holes could easily account for the observed baryon asymmetry.

We comment briefly on the potential problem that any baryon excess generated prior to the electroweak era may be erased due sphaleron
transitions~\cite{sphalerons}. The simplest possibility of avoiding sphaleron washout is to require that the black holes survive until
after the electroweak phase transition, {\it i.e.} the reheat temperature is less than the electroweak scale,
$T_{\rm evap} < T_{\rm EW} \sim 10^{-16}$.
This would imply a lower limit on the initial black hole mass,
$M_0 > 4 \times 10^{10} \left(\frac{g_*}{100}\right)^{1/6}$.  
This also forces us to a regime where $\gamma$ is very large.  Because this 
corresponds to a relatively small range of parameters, we have considered instead  the alternative of preserving the baryon asymmetry produced during black hole evaporation via $B-L$ conservation \cite{Harvey:1990qw}.  This opens up a much 
wider and more physically plausible range of masses and couplings. 
 
Assessing the final state of black hole evaporation requires a better understanding of quantum gravity. 
If black holes evaporate completely, then their final temperatures become arbitrarily large.
One might therefore worry about monopole production during the final stages of black hole evaporation. 
This was studied in \cite{Preskill:1979zi, monopoles, Krauss}, which conclude that this process is exponentially suppressed by semi-classical effects.
We also investigated the possibility that black holes form stable Planck mass relics that could act as dark matter. We found that 
this would be very constraining for black hole baryogenesis. 
To avoid relic overproduction requires large initial black hole masses $M_0 \ge 6 \times 10^{10}$. This, however, suppresses the final baryon asymmetry. 
Successful black hole baryogenesis therefore requires relatively large CP violation $\gamma$ for very small $M_X \sim 1/M_0$. 
In principle, it remains conceivable to
achieve simultaneously successful baryogenesis and a black hole relic abundance equal to the observed dark matter density, but
only if  CP violation parameter is much larger than most particle physics models predict.
The more likely possibility according to our calculations  is that relic production and black hole baryogenesis are incompatible.
Either black holes are responsible for baryogenesis, or they form stable relics which could be the dark matter.

\section*{Acknowledgments}
It is a pleasure to thank Daniel Wesley and Hitoshi Murayama for useful discussions. We are especially grateful to 
Latham Boyle and Andrew Tolley
for careful reading of the manuscript and many helpful comments. DB thanks Julian Baumann for help with the preparation of the 
manuscript. 
PJS is supported in part by US Department of Energy Grant 
DE-FG02-91ER40671.
This work was carried out while PJS was also Keck Distinguished Professor at the 
Institute for Advanced Study with support from the Wm. Keck Foundation and 
the Monell Foundation. The work of NT is supported by PPARC (UK) and the Center for Theoretical Cosmology in Cambridge.\\

\noindent
{\bf Note added:}  After completion of this paper, we received a preprint by Alexander and M\'esz\'aros \cite{Alexander:2007gj}  who considered
similar issues but based their analysis on a different black hole baryogenesis mechanism, first proposed by Nagatani \cite{Nagatani}.
Nagatani considered generating the observed net baryon asymmetry through electroweak baryogenesis,
in which a black hole with Hawking temperature above the electroweak scale creates a nearly static domain wall
around itself through which a hot plasma flows;  then, sphaleron processes within the domain wall generate the baryon asymmetry.  Alexander and M\'esz\'aros present the optimistic view that this mechanism can produce
the observed baryon asymmetry and simultaneously the observed dark matter density (assuming Planck mass black hole relics).  However, Nagatani finds that the baryon asymmetry is suppressed by a factor of $\alpha_W^5$, where $\alpha_W$ is the electroweak coupling, and the observed asymmetry can only be obtained if CP violation is maximal, ${\cal O}(1)$, within the domain wall.  This conclusion is similar to the case of the GUT baryogenesis mechanism we consider, and so does not change our qualitative interpretation that the observed baryon asymmetry and dark matter density are difficult to obtain simultaneously by the black hole evaporation mechanism.


\newpage
\begin{thebibliography}{99}
\frenchspacing

\bibitem{Cohen:1997ac}
A.~G.~Cohen, A.~De Rujula and S.~L.~Glashow,
Astrophys.\ J.\ {\bf 495}, 539 (1998)
 [arXiv:astro-ph/9707087].

\bibitem{BBN}
S.~Burles, K.~M.~Nollett and M.~S.~Turner,
Phys.\ Rev.\ D {\bf 63}, 063512 (2001)
[arXiv:astro-ph/0008495].

\bibitem{Spergel:2003cb}
D.~N.~Spergel {\it et al.},
Astrophys.\ J.\ Suppl.\  {\bf 148}, 175 (2003)  [arXiv:astro-ph/0302209];
 D.~N.~Spergel {\it et al.},
  [arXiv:astro-ph/0603449].

\bibitem{Hawking}
S.~W.~Hawking,
Nature {\bf 248}, 30 (1974); See also B.~J.~Carr, Astrophy.~J. {\bf 206}, 8 (1976).
 
\bibitem{Zeldovich}
Ya. B. Zeldovich, JETP Lett. {\bf 24}, 25 (1976).

\bibitem{earlypapers}   
D.~Toussaint, S.~B.~Treiman, F.~Wilczek and A.~Zee,
Phys.\ Rev.\ D {\bf 19}, 1036 (1979);
 A.~F.~Grillo,
Phys.\ Lett.\ B {\bf 94}, 364 (1980);
M.~S.~Turner,
Phys.\ Lett.\ B {\bf 89}, 155 (1979);
S.~W.~Hawking, I.~G.~Moss, and J.~M.~Stewart, Phys.~Rev.~D {\bf 26}, 2681 
(1982).

\bibitem{barrow}
 J.~D.~Barrow, E.~J.~Copeland, E.~W.~Kolb and A.~R.~Liddle,
 Phys.\ Rev.\ D {\bf 43}, 984 (1991).

\bibitem{therest}
A.~S.~Majumdar, P.~Das Gupta and R.~P.~Saxena,
Int.\ J.\ Mod.\ Phys.\ D {\bf 4}, 517 (1995);
E.~V.~Bugaev, M.~G.~Elbakidze and K.~V.~Konishchev,
Phys.\ Atom.\ Nucl.\  {\bf 66}, 476 (2003) [arXiv:astro-ph/0110660];
N.~Upadhyay, P.~Das Gupta and R.~P.~Saxena,
Phys.\ Rev.\ D {\bf 60}, 063513 (1999) [arXiv:astro-ph/9903253].

\bibitem{Sakharov:dj} 
A.~D.~Sakharov,
 Pisma Zh.\ Eksp.\ Teor.\ Fiz.\  {\bf 5}, 32 (1967).

\bibitem{sphalerons}
V.~A.~Kuzmin, V.~A.~Rubakov and M.~E.~Shaposhnikov,
Phys.\ Lett.\ B {\bf 155}, 36 (1985).

\bibitem{Fukugita:1986hr}
M.~Fukugita and T.~Yanagida,
Phys.\ Lett.\ B {\bf 174}, 45 (1986).

\bibitem{Harvey:1990qw}
J.~A.~Harvey and M.~S.~Turner,
Phys.\ Rev.\ D {\bf 42}, 3344 (1990).

\bibitem{inflation}
A.~H.~Guth,
Phys.\ Rev.\ D {\bf 23}, 347 (1981);
A.~D.~Linde,
Phys.\ Lett.\ B {\bf 108}, 389 (1982);
A.~Albrecht and P.~J.~Steinhardt,
Phys.\ Rev.\ Lett.\  {\bf 48}, 1220 (1982).

\bibitem{justin}
J.~Khoury, B.~A.~Ovrut, P.~J.~Steinhardt and N.~Turok,
Phys.\ Rev.\ D {\bf 64}, 123522 (2001) [arXiv:hep-th/0103239];
J.~Khoury, 
Princeton PhD Thesis, (2002).

\bibitem{cyclic}
P.~J.~Steinhardt and N.~Turok,
Science {\bf 296}, 1436 (2002);
P.~J.~Steinhardt and N.~Turok,
Phys.\ Rev.\ D {\bf 65}, 126003 (2002) [arXiv:hep-th/0111098].


\bibitem{MacGibbon}
J.~H.~MacGibbon, Nature {\bf 320}, 308 (1987).

\bibitem{Carr:1994ar}
B.~J.~Carr, J.~H.~Gilbert and J.~E.~Lidsey,
Phys.\ Rev.\ D {\bf 50}, 4853 (1994).

\bibitem{Barrow:1992hq}
J.~D.~Barrow, E.~J.~Copeland and A.~R.~Liddle,
Phys.\ Rev.\ D {\bf 46}, 645 (1992).

\bibitem{Carr:1974nx}
B.~J.~Carr and S.~W.~Hawking,
Mon.\ Not.\ Roy.\ Astron.\ Soc.\  {\bf 168}, 399 (1974).

\bibitem{blue}
B.~J.~Carr,
Astrophys.\ J.\  {\bf 201}, 1 (1975).

\bibitem{Garcia-Bellido:1996qt}
J.~Garcia-Bellido, A.~D.~Linde and D.~Wands,
Phys.\ Rev.\ D {\bf 54}, 6040 (1996) [arXiv:astro-ph/9605094].

\bibitem{Hsu}
S.~D.~H.~Hsu,
Phys.\ Lett.\ B {\bf 251}, 343 (1990).

\bibitem{extended}
D.~La and P.~J.~Steinhardt,
Phys.\ Rev.\ Lett.\  {\bf 62}, 376 (1989)
[Erratum-ibid.\  {\bf 62}, 1066 (1989)];
D.~La and P.~J.~Steinhardt,
Phys.\ Lett.\ B {\bf 220}, 375 (1989);
D.~La, P.~J.~Steinhardt and E.~W.~Bertschinger,
Phys.\ Lett.\ B {\bf 231}, 231 (1989);
E.~J.~Weinberg,
Phys.\ Rev.\ D {\bf 40}, 3950 (1989);
P.~J.~Steinhardt and F.~S.~Accetta,
Phys.\ Rev.\ Lett.\  {\bf 64}, 2740 (1990);
F.~S.~Accetta and J.~J.~Trester,
Phys.\ Rev.\ D {\bf 39}, 2854 (1989);
R.~Holman, E.~W.~Kolb and Y.~Wang,
Phys.\ Rev.\ Lett.\  {\bf 65}, 17 (1990).

\bibitem{Hawking:ga}
S.~W.~Hawking, I.~G.~Moss and J.~M.~Stewart,
Phys.\ Rev.\ D {\bf 26}, 2681 (1982).

\bibitem{Khoury:2003rt}
  J.~Khoury, P.~J.~Steinhardt and N.~Turok,
  Phys.\ Rev.\ Lett.\  {\bf 92}, 031302 (2004)
  [arXiv:hep-th/0307132].

\bibitem{Hawking:sw}
S.~W.~Hawking,
Commun.\ Math.\ Phys.\  {\bf 43}, 199 (1975).

\bibitem{Peacock:ye}
J.~A.~Peacock,
{\it Cosmological Physics}, (Cambridge Univ. Press, Cambridge, 1999).

\bibitem{Hamaguchi:2001gw}
K.~Hamaguchi, H.~Murayama and T.~Yanagida,
Phys.\ Rev.\ D {\bf 65}, 043512 (2002) [arXiv:hep-ph/0109030].

\bibitem{joel}
J.~K.~Erickson, D.~H.~Wesley, P.~J.~Steinhardt and N.~Turok,
Phys.\ Rev.\ D {\bf 69}, 063514 (2004) [arXiv:hep-th/0312009].

\bibitem{Horowitz:2002mw}
G.~T.~Horowitz and J.~Polchinski,
Phys.\ Rev.\ D {\bf 66}, 103512 (2002)
 [arXiv:hep-th/0206228].

\bibitem{Preskill:1979zi}
J.~P.~Preskill,
Phys.\ Rev.\ Lett.\  {\bf 43}, 1365 (1979).

\bibitem{monopoles}
A.~K.~Drukier and S.~Nussinov,
Phys.\ Rev.\ Lett.\  {\bf 49}, 102 (1982).

\bibitem{Krauss}
Lawrence Krauss private communication. See also
L.~M.~Krauss,
Phys.\ Rev.\ Lett.\ {\bf 49}, 1459 (1982);
L.~M.~Krauss, 
MIT PhD Thesis, (1982).

\bibitem{Chen}
P.~Chen,
[arXiv:astro-ph/0305025];
P.~Chen,
[arXiv:astro-ph/0303349];
P.~Chen,
[arXiv:astro-ph/0406514].

\bibitem{relics}
C.~G.~Callan, R.~C.~Myers and M.~J.~Perry,
Nucl.\ Phys.\ B {\bf 311}, 673 (1989);
S.~R.~Coleman, J.~Preskill and F.~Wilczek,
Mod.\ Phys.\ Lett.\ A {\bf 6}, 1631 (1991);
M.~J.~Bowick, S.~B.~Giddings, J.~A.~Harvey, G.~T.~Horowitz and A.~Strominger,
Phys.\ Rev.\ Lett.\  {\bf 61}, 2823 (1988);
K.~M.~Lee, V.~P.~Nair and E.~J.~Weinberg,
Phys.\ Rev.\ Lett.\  {\bf 68}, 1100 (1992);
G.~W.~Gibbons and K.~i.~Maeda,
Nucl.\ Phys.\ B {\bf 298}, 741 (1988);
P.~Chen,
[arXiv:astro-ph/0305025].

\bibitem{Alexander:2007gj}
  S.~Alexander and P.~Meszaros,
  [arXiv:hep-th/0703070].

\bibitem{Nagatani}
  Y.~Nagatani,
  Phys.\ Rev.\  D {\bf 59}, 041301 (1999)
  [arXiv:hep-ph/9811485];
Y.~Nagatani,
  arXiv:hep-ph/0104160.

\end{thebibliography}
\end{document}